\documentclass[%
 reprint,
superscriptaddress,
 amsmath,amssymb,
 aps
]{revtex4-2}

\usepackage{float}
\usepackage{xcolor}
\usepackage[normalem]{ulem}
\usepackage{comment}
\newcommand{\orcid}[1]{\href{https://orcid.org/#1}{\includegraphics[width=8pt]{figures/orcid.png}}}
\usepackage[colorlinks,urlcolor=goodgreen,citecolor=blue,linkcolor=goodred]{hyperref}
\definecolor{orange}{rgb}{1,0.5,0}
\definecolor{goodgreen}{rgb}{0.1,0.5}
\usepackage{graphicx}
\usepackage[dvipsnames]{xcolor}
\usepackage{dcolumn}
\usepackage{bm}
\definecolor{goodred}{rgb}{0.7,0,0}
\usepackage{mathtools}
\usepackage{pgf}

\begin{document}
\title{Coherent subgap transport in spin-split Josephson junctions}

\author{David Caldevilla-Asenjo}
\thanks{These authors contributed equally to this work}
\affiliation{Centro de Fisica de Materiales, CSIC-UPV/EHU, 20018 Donostia-San Sebastian, Spain}

\author{Gorm Ole Steffensen}
\thanks{These authors contributed equally to this work}
\affiliation{Instituto de Ciencia de Materiales de Madrid (ICMM), CSIC, 28049 Madrid, Spain}
\author{Sara Catalano}
\affiliation{Centro de Fisica de Materiales, CSIC-UPV/EHU, 20018 Donostia-San Sebastian, Spain}
\affiliation{IKERBASQUE, Basque Foundation for Science, 48009 Bilbao, Basque Country, Spain}
\author{Alberto Hijano}
\affiliation{Centro de Fisica de Materiales, CSIC-UPV/EHU, 20018 Donostia-San Sebastian, Spain}
\affiliation{Department of Physics and Nanoscience Center, University of Jyväskylä, P.O. Box 35 (YFL), FI-40014 University of Jyväskylä, Finland}
\author{Maxim Ilyn}
\affiliation{Centro de Fisica de Materiales, CSIC-UPV/EHU, 20018 Donostia-San Sebastian, Spain}
\email{maxim.ilin@csic.es}

\author{Celia Rogero}
\affiliation{Centro de Fisica de Materiales, CSIC-UPV/EHU, 20018 Donostia-San Sebastian, Spain}

\author{Ramon Aguado}
\affiliation{Instituto de Ciencia de Materiales de Madrid (ICMM), CSIC, 28049 Madrid, Spain}

\author{F. Sebastian Bergeret}
\email{fs.bergeret@csic.es}
\affiliation{Centro de Fisica de Materiales, CSIC-UPV/EHU, 20018 Donostia-San Sebastian, Spain}
\affiliation{Donostia International Physics Center (DIPC), 20018 Donostia-San Sebastián, Spain}

\author{Alfredo Levy Yeyati}
\email{a.l.yeyati@uam.es}
\affiliation{Departamento de Física Teórica de la Materia Condensada, Instituto Nicolás Cabrera and Condensed Matter Physics Center (IFIMAC), Universidad Autónoma de Madrid, Madrid 28049, Spain}

\begin{abstract}

{ We report the first experimental observation of subgap transport in  ferromagnetic insulator–superconductor–insulator–superconductor  junctions realized in EuS/Al/AlOx/Al vertical stacks. Differential conductance measurements reveal multiple Andreev reflection  peaks, with odd-order peaks split by the spin-splitting induced in the superconductor adjacent to EuS, while even-order peaks remain unaffected. Combining experiments with quasiclassical transport modeling, we extract the spin-splitting and the distribution of transmission channels, finding that a significant fraction ($\sim 23\%$) of highly transparent channels ($\tau \approx 0.9$) dominates transport. The observation of a Josephson current further confirms strong superconducting coupling through these channels. Our results demonstrate that a single spin-split superconductor is sufficient to observe the even–odd MAR effect. Our work establishes EuS/Al junctions as a versatile platform to study subgap transport, Josephson coupling, and spin-polarized superconducting phenomena.
}

\end{abstract}
\maketitle
Recent experimental advances now provide unprecedented control over subgap Andreev bound states (ABS) \cite{Sauls:PTRSA18,Martin-Rodero-Levy-Yeyati2011} in hybrid Josephson junctions based on superconductor–semiconductor devices. Arguably, this progress is mainly driven by the potential of these states for quantum technologies~\cite{AguadoAPL2020,Souto_chapter2024}, particularly qubits based on ABSs ~\cite{Janvier2015,Hays2021,PitaVidal2023,PitaVidal2024,Cheung2024} and ultimately the pursuit of topological qubits based on Majorana zero modes \cite{Nayak2008,Sarma2015,Aguado-KouwenhovenPT2020}.

Beyond these goals, the engineering of ABSs is also central to the broader field of superconducting spintronics. Here, junctions combining superconductors with magnetic materials are prime candidates for novel devices~\cite{bergeret_odd_2005,
linder_superconducting_2015,eschrig_spin-polarized_2015}. A key requirement for such functionality is lifting the spin degeneracy of the ABSs. Although this can be achieved with an external magnetic field, interfacing the superconductor (S) with a ferromagnetic insulator (F) offers a more integrated and practical solution. The F layer induces a spin-split density of states (DoS) directly, providing a significant advantage for device applications ~\cite{moodera_electron-spin_1988,BergeretRevModPhys2018}.

A prominent platform for achieving such a spin-split DoS is the EuS/Al heterostructure. In these systems, the exchange field from the ferromagnetic insulator EuS induces a substantial spin splitting in the superconducting Al layer. This is well-established in the tunneling regime, where it manifests as a characteristic splitting in the differential conductance of EuS/Al/AlOx/Al junctions \cite{meservey_spin-polarized_1994,moodera_electron-spin_1988,strambini_revealing_2017,de_simoni_toward_2018}. However, the spectroscopic properties of these spin-split junctions beyond the tunneling limit —a crucial regime for complex device operation— remain largely unexplored experimentally.
On the theoretical side, subgap transport in FI/S junctions, comprising two spin-split superconductors separated by a barrier of arbitrary transparency, has been predicted to exhibit an intriguing even–odd effect~\cite{lu_spin-polarized_2020}. 
Specifically, in symmetric junctions with noncollinear orientations of the interfacial magnetic moments, the peaks corresponding to odd-order multiple Andreev reflections (MAR) are expected to split, while those of even order remain unaffected.
Here, we put this prediction to the test by reporting a comprehensive study that elucidates the subgap transport mechanisms in spin-split superconducting junctions through a combination of experiment and theory. Our experimental study of EuS/Al/AlO$_x$/Al
hybrid junctions with transparent barriers reveals a pronounced subgap conductance that extends beyond the conventional tunneling regime.

Our theoretical modeling reveals that the characteristic even-odd effect can be generated by a single spin-split superconductor. This finding significantly simplifies the conditions for its observation, which previously required a more complex two-electrode configuration~\cite{lu_spin-polarized_2020}. The high density of transmissive channels, confirmed by our analysis, is the key reason we observe the even-odd effect, as it provides the necessary conditions for multiple Andreev reflections.

In particular, the MAR resonances in a FI/S/I/S junction appear at voltages (see Fig.~\ref{fig:3} and appendix~\ref{sec:current}):
\begin{equation}\label{eq:MAR_Energies}
 eV = 
 \begin{cases}
 \frac{\Delta_1+\Delta_2\pm h}{n} &\text{ for odd }n. \\
 \frac{2\Delta_1}{n}, \frac{2\Delta_2}{n} &\text{ for even }n.
 \end{cases}
\end{equation} 

 \begin{figure}[h]
    \centering
    \includegraphics[width=1.0\linewidth]{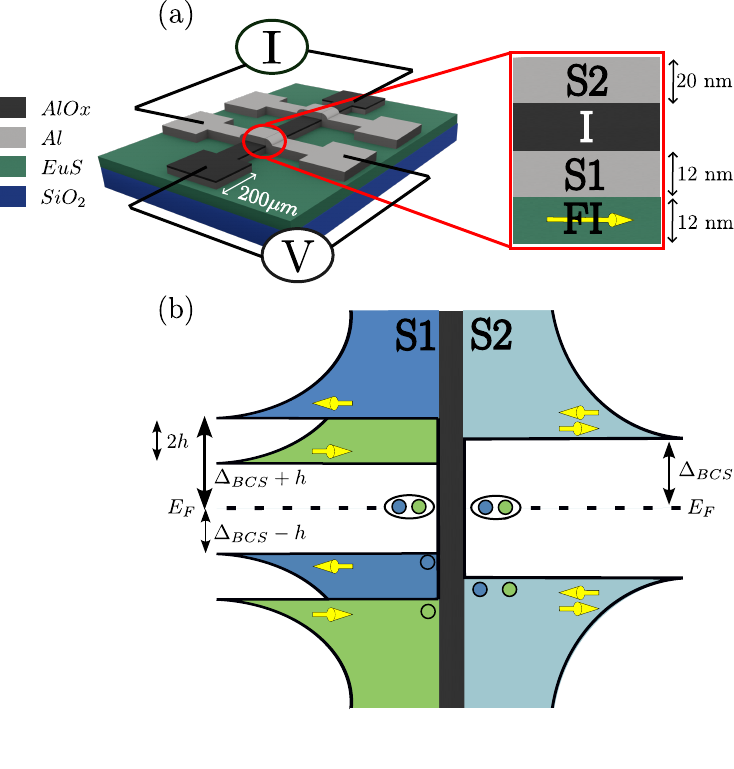}
    \caption{\label{fig:Fig1}\textbf{(a)} Device and geometry description. The $12\;\text{nm}$ bottom Al (S1) strip sits on top of the continuous $12\;\text{nm}$ EuS bottom layer (FI).  We perform tunneling spectroscopy across the oxide barrier (I) using the $20\;\text{nm}$ top Al (S2) as the second electrode. \textbf{(b)} Density of states of a BCS non-split (S2) and spin-split (S1) superconductors separated by a barrier with no applied bias voltage. The exchange interaction splits the density of states by $h$}
\end{figure}

Here, $\Delta_{1,2}$ denote the superconducting gaps of the two electrodes, and $h$ is the effective exchange field induced in one of the superconductors via the magnetic proximity effect.

\begin{figure}[h!]
    \centering
    \includegraphics[width=1.0\linewidth]{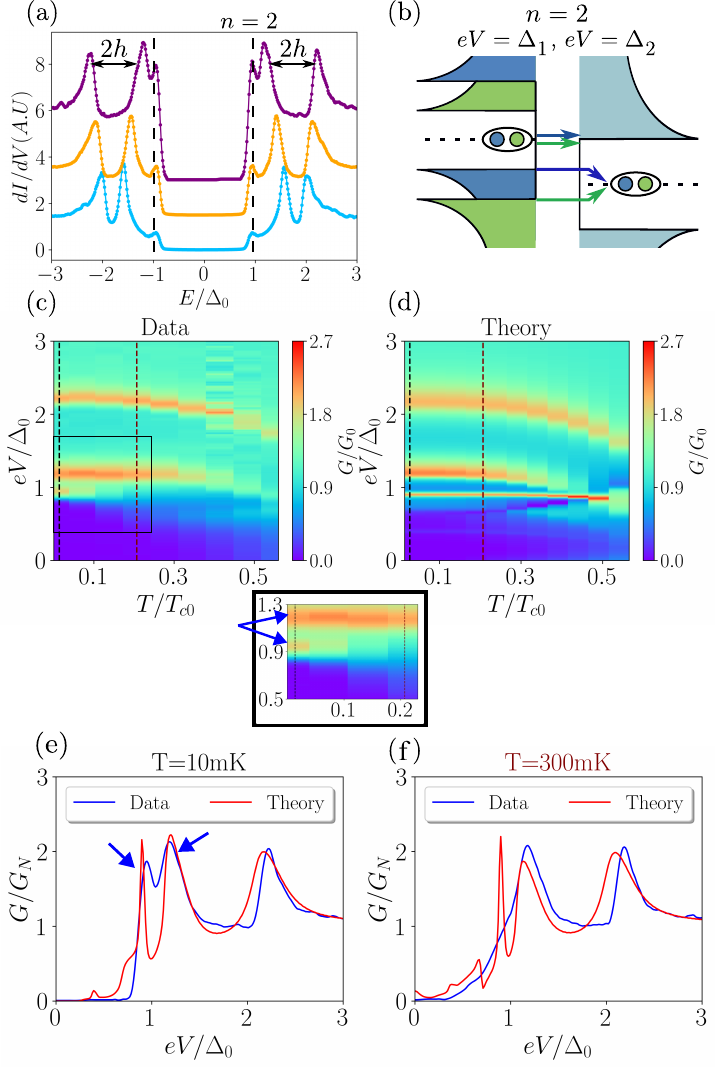}
    \caption{\label{fig:2}\textbf{(a)} Tunneling spectroscopy of three diferent devices at zero field and $10\;\text{mK}$. The curves are  displaced vertically for clarity. \textbf{(b)} Schematics of the $n=2$ MAR process, where a charge transfer of $2e$ happens between the electrodes. \textbf{(c)} [\textbf{(d)}] Temperature dependance of the experimental data [simulations] from the purple curve in panel (a). Dashed black and brown lines mark the $10$ and $300\;\text{mK}$ linecuts showed in \textbf{(e)} and \textbf{(f)}. A zoom at the black square is shown to highlight the difference between the QP peak and the MAR peak (blue arrows).\ \textbf{(e)}, \textbf{(f)} Linecut comparison between data and model at $10\;\text{mK}$ and $300\;\text{mK}$.}
\end{figure}

To verify this effect we fabricate devices schematically shown in Fig.~\ref{fig:Fig1}(a). The bottom aluminum wire is interfaced with ferromagnetic insulator EuS which creates spin-split DoS represented in Fig.~\ref{fig:Fig1}(b). Details of preparation and thorough characterization of these tunneling junctions with highly resistive tunneling barrier were published in our previous work~\cite{hijano_coexistence_2021}. Here, we optimized the transparency of the tunneling barrier to measure subharmonic gap structures and investigate the effect of the spin-split field on the MAR subgap structure. More details about the fabrication are found on the appendix~\ref{sec:Methods}.

Results of tunneling spectroscopy measurements performed for three different devices in zero external magnetic field are depicted in Fig.~\ref{fig:2}(a). All three devices show three symmetric conductance peaks with respect to $V=0$. The two  outermost peaks are the single  quasiparticle (QP) spin-split peaks from Fig.~\ref{fig:Fig1}(b), observed in previous works \cite{hijano_coexistence_2021,strambini_revealing_2017}. From the position of these peaks we obtain the strength of the spin-splitting field $h=45\; \mu eV$, $75 \; \mu eV$ and $100\;\mu eV$, which is equivalent to the applied effective magnetic fields of $0.41\;T$, $0.68\;T$ and $0.91 \;T$, respectively. These results match to the expected values of $h$ for the selected  growth conditions of EuS~\cite{geng_superconductor-ferromagnet_2023,gonzalez-orellana_exploring_2023}.

In addition, all three samples show a peak in dI/dV spectra at $E=\pm\Delta_{0}=220\;\mu\text{eV}$, marked by the vertical dashed lines in Fig.~\ref{fig:2} (a). This feature remains at the same energy independently of $h$, indicating the non-magnetic origin of this peak. We attribute it to subgap charge transport, namely to the second-order ($n=2$) Andreev reflection, which according to Eq.~\eqref{eq:MAR_Energies} appears at the bias $eV = \Delta$, provided that $\Delta_1$ = $\Delta_2$. This process, transferring a Cooper pair from one electrode to the QP states of the other, is illustrated in Fig.~\ref{fig:2}(b)~\cite{Zimmermann1995Mar, lu_spin-polarized_2020}. The near coincidence of $\Delta_1$ and $\Delta_2$, yielding a single $n=2$ peak, is consistent with the similar thicknesses of the top and bottom Al wires and the low measurement temperature~\cite{buzdin_proximity_2005}.
Temperature dependence of the  tunneling spectra, measured for the device with the highest spin-splitting field, are shown  as a color plot in Fig.~\ref{fig:2}(c). A shift of the QP peaks towards smaller energies is clearly seen. This is  caused by the decrease  of the order parameter $\Delta$ when the temperature approaches the critical value $T_{c0}=1.45K$. Details about the $T_c$ can be found in the appendix~\ref{sec:Tc and Hc}.  Zooming into the region close to $eV=\Delta_0$ shows that the peak attributed to the n=2 MAR process rapidly broadens with increasing temperature. Tunneling spectra collected at 10~mK and 300~mK (Fig.~\ref{fig:2}(e) and(f)) show that it becomes no longer visible for $T>300\;\text{mK}$. 

For the theoretical description of  our results, we use the quasiclassical formalism combined with a Keldysh-Floquet method~\cite{Cuevas1996Sep} to obtain the current due to MAR processes.
Expressions are presented in the appendix~\ref{sec:quasiclassical GF} and ~\ref{sec:current}. 
The inputs in our model are: the  BCS pairing amplitude on each lead, $\Delta_{1,2}$, spin-splitting field, $h$, spin-flip relaxation time $\tau_{\mathrm{sf}}$ and the transparency for each channel, $\left\{\tau_{i}\right\}$. $\Delta_{1,2}$ and $h$ can be directly obtained from the experimental data by measuring the outer QP peak positions; while $\left\{\tau_{i}\right\}$ and $\tau_{\mathrm{sf}}$ are fitting parameters. As a secondary approach we can experimentally estimate the latter by measuring $T_{c}$ and the peak broadening (see Eq.~\ref{eq:TcFormula}, leaving the effective transparency, $\tau$, as the main fitting parameter. 
By fitting the data from Fig.~\ref{fig:2}(c) we found that $\tau_{\mathrm{sf}}=(0.12\Delta_{0})^{-1}$ fits the results well. According to Eq.~\ref{eq:TcFormula} and the measured $T_{c}$ we get an experimental value of $\tau_{\mathrm{sf}}=(0.09\Delta_{0})^{-1}$, in good agreement with the fitting. For more details see appendix~\ref{sec:Tc and Hc}. Other  parameters are $\Delta_1=\Delta_0=220\;\mu\text{eV}$, $\Delta_2=205\;\mu\text{eV}$, $h=100\;\mu\text{eV}$ and a single transmission value $\tau =0.2$ for all $N=150$  channels. $N$ is obtained by normalizing to the normal conductance of the junction, $G(V\gg \Delta_1+\Delta_2)$. The results of these simulations are shown in Fig.~\ref{fig:2}(d), combined with a direct comparison between the measured (blue) and simulated (red) tunneling spectra, presented in Fig.~\ref{fig:2}(e) and (f) for $T=10\;\text{mK}$ and $T=300\;\text{mK}$, respectively. The experimental features are closely reproduced by the simulations and are consistent with our hypothesis about the origin of the third peak being the coherent Cooper-pair tunneling at the $n=2$ MAR. Compared to the experiments, the second order MAR in the simulations exhibits a weaker thermal broadening suggesting some broadening effects not captured by our model [linecut at $300\;\text{mK}$ in panel (f)]. Also, the simulations at $10\;\text{mK}$ present a richer subgap structure, not present in experiment, indicating further broadening due to higher order processes. Additionally, at $T\approx 0.3T_{c0}$ supgap features appear in both theory and experiment, related to thermally activated processes at $eV = h \pm |\Delta_1-\Delta_2|$, due to filling of the QP states \cite{meservey_spin-polarized_1994}. 

\begin{figure}[h]
    \centering
    \includegraphics[width=1.0\linewidth]{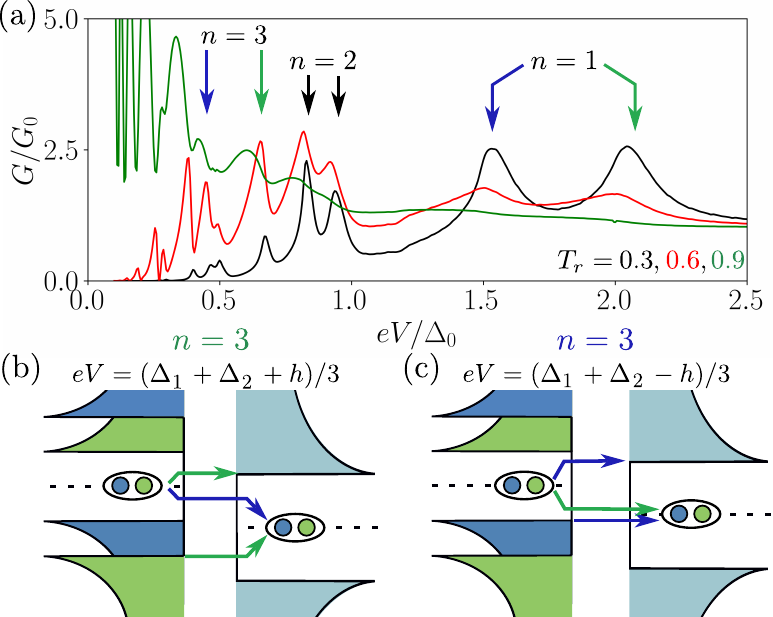}
    \caption{\label{fig:3} Theoretical predictions of the higher MAR orders.\textbf{(a)} Numerical simulation of the tunneling conductance as a function of the barrier transparency for $\tau=0.3$, $0.6$ and $0.9$. The rest of the parameters are fixed at $\Delta_1=220\;\mu\text{eV}$, $\Delta_2=190\;\mu\text{eV}$, $h=55\;\mu\text{eV}$ and $\tau_{\mathrm{sf}}=(0.02\Delta_{0})^{-1}$. Higher transmission resolve higher MAR orders while broadening the outer features. \textbf{(b)} and \textbf{(c)} Schematics of a $n=3$ MAR process for the two spin selection possibilities (blue, green). This odd order transfers three charges which creates the spin-split resonance. } 
\end{figure}

\begin{figure*}[t]
    \centering
    \includegraphics[width= 0.9\linewidth]{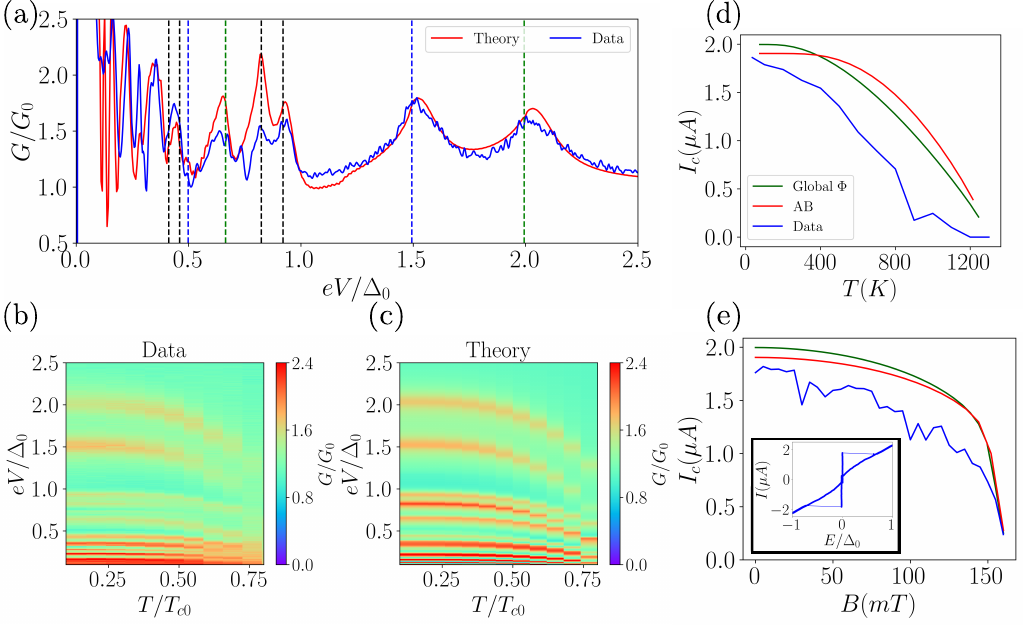}
    \caption{\label{fig:4} Full characterization of the device with lower interface resistance. \textbf{(a)} Tunneling conductance (blue) and simulations (red) at $30\;\text{mK}$. 
    Vertical line cuts show the  different MAR orders, odd spin-split (blue-green) and even not spin-split (black). \textbf{(b)} [\textbf{(c)}] Temperature dependence of the tunneling conductance from the experiment [simulations]. Features decay to smaller energies as $T \to T_{c}$. \textbf{(d)} and \textbf{(e)} Josephson critical current measured (blue) from the IV curves (inset in (e)), modified Ambegaokar-Baratoff prediction (Eq.~\ref{eq:AGIc}, red) and going beyond the tunneling limit  (Eq.~\ref{eq:FullIc}, green) against the temperature and magnetic field of the system. The inset in panel \textbf{(e)} shows the measured IV around 0. $I_c=\text{max}(I[E=0])$}
\end{figure*}

 The still low transparency of the barrier and the experimental broadening prevent higher MAR orders to be resolved in these devices. Simulations of conductance characteristics for higher transparencies within our model are shown in Fig.~\ref{fig:3}(a). For higher transmission probability $\left(  \tau > 0.3 \right)$, more subgap peaks emerge,  corresponding to higher MAR processes, while the outer peaks are broadened. As discussed, the presence of the FI interface in our system creates an energy difference for each spin selection according to Eq.~\eqref{eq:MAR_Energies} when an odd number of single-charges are transferred. The $n=3$ process, compared to the $n=2$ case, involves an extra Andreev reflection, allowing an extra single-charge to transfer across the junction. $h$ creates an energy difference between the two spin species, resulting on two resonances for the n=3 case, as depicted in Fig. \ref{fig:3}(b),(c). The two peaks found for $n=2$ are due to $\Delta_1\neq \Delta_2$

 To validate these predictions, we measure a device with lower barrier resistance. In Fig.~\ref{fig:4}(a) we show the experimental conductance spectrum (blue) of a device with richer subgap structure, featuring several peaks. Simulations (red) demonstrate that such peak positions are consistent with MAR resonances, highlighted by vertical dashed lines. Indeed, the even-order MAR (vertical black lines) are split, proving that in this device we can resolve $\Delta_1 - \Delta_2$. Second, the odd-order MAR (blue-green vertical lines) reflects the spin-splitting of the bottom electrode. The data is well reproduced by the simulations  considering the same parameters as in Fig. \ref{fig:3}.  Instead of a single $\tau$, for this junction we considered a distribution of transmissions, $\tau_{i}$, for different channels. A good agreement is found for the total transmission channels $N=150$ distributed as $N_{\tau=0.3}=77$,\;$N_{\tau=0.6}=39$ and $N_{\tau=0.9}=34$. Although the junction area is very large, $0.04\;mm^2$, the relatively low number of channels suggests that transport occurs mainly where the barrier is thinner, through the high transparency channels, greatly reducing the effective junction area.  We also present the temperature dependence of the differential conductance in Fig.~\ref{fig:4}(b) and the theoretical simulation in Fig.~\ref{fig:4}(c),  from the base $T$ to $T_{c}$ of the system. Again, the $\Delta(T)$-like dependence is appreciated by the main features, as the resonant energy decays as the temperature increases up to $T_{c}$. Experiment and theory are in very good agreement, showing that the broadening effects present in Fig.~\ref{fig:2} are not relevant in this sample. 
 
The enhanced transparency not only affects the MAR features, it also results in a sizable Josephson supercurrent at $V=0$. This is another key result as it confirms that phase-coherent Cooper pairs transport survives the magnetic interaction in this system. The ability to sustain a supercurrent in FI/S/I/S systems is crucial  for developing spintronic and hybrid quantum devices that aim to combine magnetic interaction with coherent transport. 
The inset in Fig.~\ref{fig:4}(e)  shows an I-V characteristic of the device, exhibiting a Josephson critical current, $I_c$,  of $1.7\;\mu A$ Theoretically, we have computed  $I_c$ using  two approaches. First we modified the Ambegaokar-Baratoff (AB) relation \cite{ambegaokar_tunneling_1963} by including spin-split terms and neglecting the highest transparent channels, which allowed us to simulate $I_{c}$ by taking the normal resistance $R_{N}$ (see Eq.~\ref{eq:AB} in appendix~\ref{sec:current}). The result is shown by the red curves in Fig.~\ref{fig:4}(d-e)). Secondly, taking the channel distribution from Fig.~\ref{fig:4}(a), we calculated the $I_c$ by considering a constant phase drop across the junction which maximizes $I_c$ (see Eq.~\ref{eq:FullIc}). This results in the green curves in Fig.~\ref{fig:4}(d-e)). The measured value of the switching current of $I_{c}\approx 1.7 \;\mu A$ is in good agreement with the expected value of $1.9\; \mu A$ predicted by the AB and $2 \;\mu\text{A}$ predicted by the second method . We compare the experimental $I_{c}$ data (blue curve, Fig.~\ref{fig:4}(d) and (e)) with the temperature (d) and in-plane magnetic field (e) predictions from both models and similar behaviour is found, further supporting the agreement between theory and experiment. We note here the lack of any Fraunhofer pattern in Fig.~\ref{fig:4}(e). This further suggests that transport occurs only through highly transparent channels, making the effective area much smaller. As a result, magnetic flux is strongly quenched, requiring fields larger than the critical field to introduce a single flux quantum through the junction.
Further investigation in this direction, such as structural characterization of oxide defects across the junction area, is needed but is beyond the scope of the present work.

In summary, we report the first experimental observation of subgap transport in tunneling junctions with a spin-split superconductor. Combining experiment and theory, we identify even and odd MAR processes, with the odd-order peaks affected by the spin-splitting field. In low-resistance junctions, a significant fraction of highly conductive channels ($\sim 23\%$) account for the observed spin-split MAR, and we also measure a Josephson current coexisting with the spin splitting. Our results provide a robust platform to study subgap transport and Josephson coupling mediated by spin-split superconductors, which can be extended to explore FI/S/I/S junctions, Josephson valves~\cite{bergeret_enhancement_2001}, unconventional $\pi$-junctions, and spin-polarized supercurrents involving both singlet and triplet condensate components~\cite{bergeret_electronic_2012}.

{\it Acknowledgments}
We thank financial support from the European Union’s Horizon Europe research and innovation program under grant agreement No. 101130224 (JOSEPHINE) and from the Spanish MCIN/AEI/10.13039/501100011033 
through the grants PID2023-148225NB-C31, TED2021-130292B-C41, PID2024-160189NA-I00, TED2021-130292B-C42 and PID2020-114252GB-I00

\appendix
\section{Experimental methods}\label{sec:Methods}

We fabricated vertical $\text{Al}(12\;\text{nm})/\text{AlOx}/\text{Al}(20\;\text{nm}$) junctions in which the bottom Al electrode is grown on top of a $12\;\text{nm}$ thin film of EuS. An illustration of a typical sample, consisting of two Al/$AlOx$/Al junctions, is depicted in Fig.~\ref{fig:Fig1}.  EuS is a ferromagnetic insulator with in plane magnetization, small coercive field (varying from few to tens of mT,~\cite{PhysRevLett.110.097001,strambini_revealing_2017, hijano_coexistence_2021} ) and a bulk Curie temperature of $16.7\;K$ \citep{mauger_magnetic_1986}, however in the thin film regime this can increase up to $20\;K$ \cite{aguilar-pujol_magnon_2023}. This material preserves its magnetic properties even if the growth is not epitaxial \cite{aguilar-pujol_magnon_2023}. Combined Al/EuS interfaces deposited \textit{in-situ} have previously shown to be stable, free from chemical defects \cite{gonzalez-orellana_exploring_2023}.

EuS is grown as a continuous film of $12\;\text{nm}$ thickness on top of $SiOx$ $100\;\text{nm}$. The sublimation temperature employed during the evaporation of EuS affects the exchange interaction $h$ \cite{gonzalez-orellana_exploring_2023}, sometimes even leading to a full supression of the superconductivity \cite{hijano_coexistence_2021}. It is important to operate this growth on the weak-interaction range so $h$ and the Al superconductivity can coexist. Al is evaporated through a hard mask defining a stripe of $200\;\mu m$ width on the substrate.  $AlOx$ is induced by exposing the deposited Al wire onto a low pressure atomic $O$ environment originated with a non-accelerated microwave plasma source. This low-energy procedure works as a stochastic probability event that preserves Al composition and sometimes can lead to pinhole formation through the barrier, resulting in the measured subgap structure~\cite{PhysRevLett_72_1738}. After the oxidation a second hard mask is used to evaporate the top $20\;\text{nm}$ Al strips. More details about the fabrication process can be found in Ref~\cite{hijano_coexistence_2021,gonzalez-orellana_exploring_2023}

As illustrated in Fig.~\ref{fig:Fig1}, each sample consists of two junctions. The I-V of the junctions are measured in 4-probe configuration, as sketched. The dI/dV curves are then obtained by numerical differentiation of the measured I-V characteristics. The samples are characterised in a dilution refrigerator with a base temperature of $10\;\text{mK}$. Current and voltage signals are sourced and measured with two Keithley 2450 Source Measure Unit (SMUs) and a Standford Research SR560 voltage pre-amplifier.

\section{Quasiclassical Green's function formalism}\label{sec:quasiclassical GF}
In our model  we use  the quasiclassical Green's function (GF) formalism~\cite{Eilenberger,Larkin-Ovchinnikov:1968,Usadel,volkov1993proximity,lambert1998phase,belzig1999quasiclassical}  to compute the differential conductance of the junction. Since we are dealing with non-equilibrium properties of superconductors with spin-dependent fields, the quasiclassical Green's function $\bar{g}$ is a $8 \times 8$ matrix in Keldysh-Nambu-spin space.

In dirty materials, the quasiclassical equations reduce to a diffusive-like equation, known as the Usadel equation~\cite{Usadel}. We assume that the superconducting electrodes are homogeneous, this assumption is justified if the thickness of the superconductors is much smaller than the superconducting coherence length \cite{hijano_coexistence_2021,strambini_revealing_2017}. In this case  the homogeneous superconductor is described by the Usadel equation
\begin{equation}\label{usadel0}
    [i\epsilon\hat{\tau}_3+\Delta\hat{\tau}_2-i\boldsymbol{h}\cdot\boldsymbol{\sigma}\hat{\tau}_3-\bar{\Sigma}_{\mathrm{sf}},\check{g}]=0\; ,
\end{equation}
and the normalization condition $\check g^2=1$. Here $D$ is the diffusion constant, $\epsilon$ is the energy, $\Delta$ is the superconducting order parameter, $\boldsymbol{h}$ is the effective exchange field induced in the superconductor via the magnetic proximity effect, and $\bar{\Sigma}_{\mathrm{sf}}$ is the self-energy term describing spin-flip relaxation processes
\begin{equation}\label{sf relaxation}
    \bar{\Sigma}_{\mathrm{sf}}=\frac{\hat{\sigma}_i\hat{\tau}_3\check{g}\hat{\tau}_3\hat{\sigma}_i}{8\tau_{\mathrm{sf}}}\; ,
\end{equation}
with $\tau_{\mathrm{sf}}$ the spin-flip relaxation time.  The sum over repeated indices is implied. 
The experiments in the present work are made using Al layers, for which the spin-orbit scattering can be neglected. The matrices  $\hat{\sigma}_i$ and $\hat{\tau}_i$ ($i=1,2,3$) in Eq.~\eqref{usadel0}  are the Pauli matrices in spin and Nambu space, respectively.

 In addition, the order parameter needs to be computed self-consistently, where the self-consistency equation is~\cite{Kopnin:2001}:  
\begin{equation}\label{self-consistency0}
    \Delta\ln{\frac{T_{c0}}{T}}=2\pi T\sum_{\omega_n>0}\left(\frac{i}{2}\mathrm{Tr} \hat{f}+\frac{\Delta}{\omega_n}\right)\; ,
\end{equation}
where $T_{c0}$ is the field-free critical temperature and $\hat{f}$ is the anomalous part of the Matsubara GF $\check{g}=\hat{g}\hat{\tau}_3 +i\hat{f}\hat{\tau}_2$, obtained by analytical continuation $\epsilon \rightarrow i\omega_n$, where $\omega_n=2\pi T(n+1/2)$, $n \in \mathbb{Z}$, are the Matsubara frequencies~\cite{Abrikosov-Gorkov-Dzyaloshinski}. From this formula, we obtain a self-consistency equation for the critical temperature, $T_c$, by expanding $\hat{f}$ in small $\Delta$,
\begin{equation}
\ln\left(\frac{T_c}{T_{c0}}\right)=\psi\left(\frac{1}{2}\right) - \text{Re }\psi\left(\frac{1}{2}+\frac{1}{2\pi T_c\tau_{sf}}+i\frac{|\boldsymbol{h}|}{2\pi T_c}\right), \label{eq:TcFormula}
\end{equation}
with $\psi(x)$ denoting the digamma function.
The Usadel equation, together with the normalization condition and the self-consistently equation determine the value of the GF, from which non-equilibrium properties such as the charge current may be computed as presented in the next section.

\section{MAR and critical current}\label{sec:current}
We consider the system to be composed of a left and right diffusive superconductor, each described by a quasiclassical Green's function, connected by $N$ transmission channels with tranmissions $\tau_i$. Following Ref.~\cite{Nazarov1999May} the matrix current contribution from of a channel $i$ is given by,
\begin{equation}
\bar{I}_i(t) = \frac{e^2}{h}\frac{\tau_i \left[\bar{g}_1,\bar{g}_2\right]_-}{1-\frac{1}{2}\tau_i +\frac{1}{4}\tau_i \left[\bar{g}_1,\bar{g}_2\right]_+}\left(t,t\right)
\end{equation}
where $-(+)$ describe (anti-)commutators and with time-convolution assumed in the matrix structure, $\bar{g}_1\bar{g}_2(t,t') = \int_{-\infty}^{\infty}dt''\bar{g}_L(t,t'')\bar{g}_R(t'',t')$. The Green's functions are written in Nambu-Keldysh space,
\begin{align}
\bar{g}_j(t,t') &= \begin{pmatrix}
\check{g}_j^R(t,t') && \check{g}_j^K(t,t') \\
0 && \check{g}_j^A(t,t') 
\end{pmatrix}, \\ 
\check{g}^R_j(t,t') &= \int_{-\infty}^\infty d\omega\hspace{0.1cm} e^{ieV_jt\hat{\tau}_3}\check{g}^R_j(\omega)e^{-ieV_it'\hat{\tau}_3}e^{-i\omega(t-t')}, \label{eq:TimeDependg}
\end{align}
where $\check{g}_j^R(\omega)$ is the solution to Eq.~\eqref{usadel0} with $\check{g}^A_j(\omega) = -\hat{\tau}_3\left[\check{g}^R_j(\omega)\right]^\dagger\hat{\tau}_3$ and $\check{g}^K_j(\omega) = \left(\check{g}^R_j(\omega)-\check{g}^A_j(\omega)\right)\tanh\left(\omega/2T\right)$. The $eV$ time-dependence in Eq.~\eqref{eq:TimeDependg} relates to the AC Josephson effect, in which a finite voltage difference $V = V_1 - V_2$ results in a time-dependent response with periodicity $T_p=2\pi/(eV)$. For simplicity we will in the following choose the gauge $V_1 = 0$ and $V_2 =V$. Next, to capture the time-dependence we assume that the system reaches a time-periodic steady state, $\bar{g}_j(t,t') = \bar{g}_j(t+T_p,t'+T_p)$ and $\bar{I}_i(t) = \bar{I}_i(t+T)$. This periodicity allows us to use Floquet theory to define,
\begin{align}
\bar{g}_{j,mm'}(\omega) = \frac{1}{T_p}&\int_{0}^{T_p} dt \hspace{0.1cm} e^{i(m-m')eVt} \\ &\times\int_{-\infty}^\infty dt'\hspace{0.1cm} e^{i(\omega+m'eV)(t-t')}\bar{g}_{j}(t,t'), \nonumber
\end{align}
where index $mm'$ relates to matrix entrances in the Floquet matrix, $\underline{g}$ such that each $\bar{g}_{i,mm'}(\omega)$ is a $8\times8$ matrix in Keldysh-Nambu space. Using these Floquet matrices, the DC component of the charge current, $I_{n}$, can be expressed from the matrix current,
\begin{equation}\label{eq:Theor_I}
I_{i} = \frac{1}{16e}\text{Tr}\hspace{0.1cm}\hat{\tau}_3 \bar{I}^K_{i,0}
\end{equation}
with the factor $16$ arising from the $4\times4$ spin-Nambu space. Superscript $K$ indicates the Keldysh component of,
\begin{equation}
\bar{I}_{i,0} = \frac{e^2}{h}\int_{-\infty}^\infty d\omega\left[\frac{\tau_i \left[\underline{g}_1,\underline{g}_2\right]_-}{1-\frac{1}{2}\tau_i +\frac{1}{4}\tau_i \left[\underline{g}_1,\underline{g}_2\right]_+}\right]_{00} \label{Seq:FullDCCurr}
\end{equation}
with assumed matrix multiplication in Floquet space, and final indices $00$ referring to the DC Floquet component. Broadly speaking, the $m$-th Floquet components captures the $m$-th order of MAR \cite{Cuevas1996Sep}, where in numerics the Floquet dimension is truncated at $m$ large enough to assure convergence of $I_{i,0}$. For more details on derivations and numerical strategies to solve this, see Ref.~\cite{Ibabe2023May} supplemental material. 
Next, we describe how critical current is obtained. First, we assume that voltage is zero, $V=0$, and instead a static phase difference is stabilized across the junction, $\phi = \phi_1 - \phi_2$, with $\phi_j$ replacing $eV_jt$ in Eq.~\eqref{eq:TimeDependg}. This removes the Floquet structure and results in a current, $I_i(\phi)$, given by Eq.~\eqref{eq:Theor_I}, with a general non-sinusoidal phase dependence due to finite transmission. To obtain an Ambegaokar-Baratoff like limit with sinusoidal phase dependence, we assume low transmission $\tau_i\ll1$ such that the full DC current is given by,
\begin{align}
I_i(\phi) &= \frac{e\tau_i}{16h}\text{Tr}\hspace{0.1cm} \hat{\tau}_3 \int_{-\infty}^\infty d\omega\hspace{0.1cm}\left[\overline{g}_1,\overline{g}_2\right]_-^K, \\
&=\frac{ie\tau_i}{2h} \sin\phi\hspace{0.1cm}\text{Tr}_\sigma \hspace{0.1cm} \int_{-\infty}^\infty d\omega \hspace{0.1cm} \text{Re}\left(\hat{f}_1^R\hat{f}_2^R\right)\tanh\left(\frac{\omega}{2T}\right).
\end{align}
From this equation, we identify the critical current as $I_c = \sum_iI_i(\pi/2)$, which in Matsubara frequencies can be expressed as,
\begin{equation}
I_c = G\frac{k_BT}{e}\text{Tr}_\sigma\sum_{\omega_n>0}\text{Re}\left(\hat{f}_1^R(\omega_n)\hat{f}_2^R(\omega_n)\right), \label{eq:AGIc}
\end{equation}
which we solve numerically to yield the values presented in Fig.~\ref{fig:4} as 'AB'. Here, $G=\frac{2e^2}{h}\sum_i\tau_i$ is the measured normal state conductance. For two standard BCS superconductors, $\hat{f}_i^R(\omega_n)=\frac{\Delta}{\sqrt{\Delta^2+\omega_n^2}}$, one finds,
\begin{equation}
I_c = \frac{\pi\Delta G}{2e}\tanh\left(\frac{2\Delta}{k_BT}\right), \label{eq:AB}
\end{equation}
which we recognize as the typical Ambegaokar-Baratoff formula. In our model, we also include effects of spin-splitting, spin-scattering, unequal gaps, and effects of temperature on $\Delta$, which results in curves that are qualitatively similar to Eq.~\eqref{eq:AB}.

In order to account for the high transparency channels we also studied a different approach. Here, we use the fitted transmission values from the MAR spectroscopy to numerically calculate $I_c$ to all orders in transmission. The critical current for all channels is then defined as, 
\begin{equation}
I_c = \text{max}_\phi \sum_i I_i(\phi), \label{eq:FullIc}
\end{equation}
which is used to obtain the red curves in Fig.~\ref{fig:4}(d) and (e) in the main text. This approach requires more information about the barrier than conductance, $G$, unlike Eq.~\eqref{eq:AGIc}, due to its explicit dependence on the transmission distribution.

\section{$T_c$ and $H_{c}$ measurements}\label{sec:Tc and Hc}

Following Eq.~\eqref{eq:TcFormula}, we can extract the experimental value of $\tau_{sf}$ by comparing the drop in the superconducting critical temperature of our Al films by the effects of the EuS. 
\begin{figure}[h]
    \centering
    \includegraphics[width=1.0\linewidth]{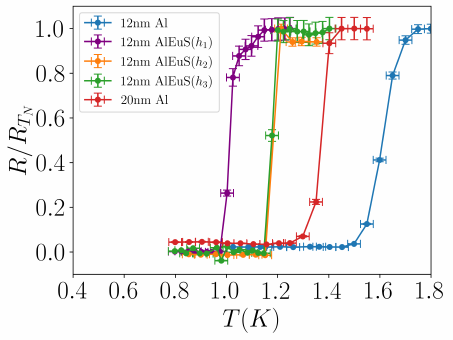}
    \caption{\label{fig:Tc_meas}\color{black}{Experimental measurement of $R(T)$ for different samples. 2 samples shown in Fig.~\ref{fig:2}(a), purple curve labeled as $h_{1}=100\;\mu\text{eV}$, and the orange curve, labeled as $h_{2}=75\;\mu\text{eV}$; and the sample shown in Fig.~\ref{fig:4}(b), labeled as $h_{3}=55\;\mu\text{eV}$. The blue curve refers to the $12\;\text{nm}$ bottom, non-proximitized Al that serves as $T_{C0}$ in Eq.~\eqref{eq:TcFormula}. The red curve is the top $20\;\text{nm}$ Al, which is supposed to be the same for every sample for simplicity. Data is normalized to the resistance on the normal state $R_{T_{N}}$}} 
\end{figure}

In Fig.~\ref{fig:Tc_meas} we present the R(T) from the presented samples. The followed criteria for $T_{c}$ is $R(T_{c})=0.1R_{T_{N}}$, being $R_{T_{N}}$ the resistance on the normal state. Using these criteria we measured for the $12\;\text{nm}$ non-proximitized Al $T_{c0}=1.45\pm0.03\;\text{K}$ (blue curve). This value serves as the reference $T_{c0}$ using Eq.~\eqref{eq:TcFormula}. For the top, $20\;\text{nm}$ Al, we found $T_{c}=1.28\pm0.3\;\text{K}$ (red curve), which is taken as constant at every sample for simplicity. Samples showed in Fig.~\ref{fig:2}(a) (purple curve labeled as $h_1=100\;\mu\text{eV}$ and orange curve labeled as $h_2=75\;\mu\text{eV}$) and Fig.~\ref{fig:4}(b) (green curve labeled as $h_3=55\;\mu\text{eV}$) are measured, showing respectively, $T_{c,h_{1}}=0.98\pm0.01$ and $T_{c,h_{2}}\approx T_{c,h_{3}}=1.15\pm0.01$. In the following table we present the values of $\tau_{sf}$ obtained from~\ref{eq:TcFormula} using the experimental results (\textbf{$\tau_{sf}\;\text{Eq.}$}) and those obtained from the fittings (\textbf{$\tau_{sf}\;\text{fit}$}):
\begin{table}[H]
\centering
\caption{\label{T:tau_sf}Comparison between $\tau_{sf}$ values eq and fittings}
\begin{tabular}{|c|c|c|} 
\hline
\textbf{Al electrode}                 & \textbf{$\tau_{sf}\;\text{Eq.}$} & \textbf{$\tau_{sf}\;\text{fit}$}   \\
\hline
$12\;\text{nm},\;h_1=100\;\mu\text{eV}$ & $1/(0.09\Delta_{0})$    & $1/(0.12\Delta_{0})$   \\
$12\;\text{nm},\;h_2=75\;\mu\text{eV}$  & $1/(0.075\Delta_{0})$   & $1/(0.075\Delta_{0})$  \\
$12\;\text{nm},\;h_3=55\;\mu\text{eV}$  & $1/(0.1\Delta_{0})$     & $1/(0.02\Delta_{0})$   \\
\hline
\end{tabular}
\end{table}

From the Table~\ref{T:tau_sf} we can see that comparing between data and theory values match within a $\sim 15\%$ discrepancy. These results demonstrate a good agreement between data and the model used for describing the results.

\begin{figure}[h]
    \centering
    \includegraphics[width=1.0\linewidth]{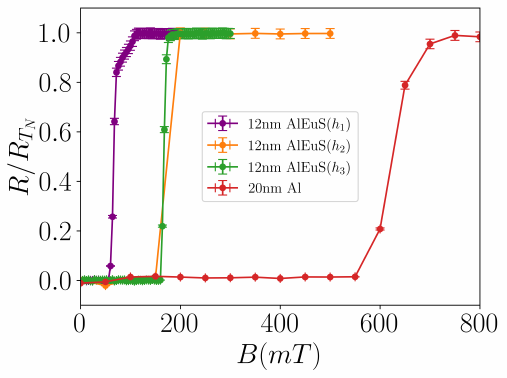}
    \caption{\label{fig:Hc_meas}\color{black}{Experimental data of $H_{c}$ for the same samples as in Fig.~\ref{fig:Tc_meas}. the Al $12\;\text{nm}$ sample was not measured.}}
\end{figure}

We also present the data corresponding to the in-plane critical magnetic field $H_{c}$ for the same samples. Using the same criteria as $T_{c}$ on the curves depicted in Fig. \ref{fig:Hc_meas} gives $H_{c,h_{1}}=60\pm2\;\text{mT}$, $H_{c,h_{2}}=150\pm25\;\text{mT}$, $H_{c,h_{3}}=160\pm2\;\text{mT}$ and $H_{Al\;20\text{nm}}=550\pm25\;\text{mT}$. If we compare these values with the applied $h$, we see that for $h_{1}$ and $h_{2}$, there is a difference of $\approx 90\;\text{mT}$, which is similar to the expected difference among the magnetic couplings $h_{1}-h_{2}=130\;\text{mT}$. As in Fig.~\ref{fig:Tc_meas}, sample $h_{3}$ shows a fairly similar $H_{c}$ to $h_{2}$ while having roughly a $75\%$ matching exchange interaction $h$. These results show that a higher $h$ reduces the $H_{c}$ due to the stronger magnetic properties of the electrode.

\bibliography{biblio}
\end{document}